\documentclass[prl,twocolumn,floatfix,showpacs]{revtex4}
\usepackage{graphics}
\usepackage{dcolumn}
\usepackage{bm}
\usepackage{epsfig}
\usepackage{pictex}

\begin{document}
\draft
\preprint{HEP/123-qed}
\title{New possibility of the ground state of quarter-filled one-dimensional strongly correlated electronic system interacting with localized spins}
\author{Chisa Hotta}
\address{Aoyama-Gakuin University, 5-10-1, Fuchinobe, Sagamihara, Kanagawa 229-8558}
\author{Masao Ogata}
\address{Department of Physics, University of Tokyo, Hongo, Bunkyo-ku, 
Tokyo 113-0033}
\author{Hidetoshi Fukuyama}
\address{International Frontier Center for Advanced Materials, IMR, Tohoku University, Miyagi 980-8577}
\date{\today}
\begin{abstract}
We study numerically the ground state properties of the one-dimensional 
quarter-filled strongly correlated electronic system interacting antiferromagnetically 
with localized $S=1/2$ spins. 
It is shown that the charge-ordered state is significantly 
stabilized by the introduction of relatively small coupling with the localized spins. 
When the coupling becomes large the spin and charge degrees of freedom behave quite independently 
and the ferromagnetism is realized. Moreover, the coexistence of ferromagnetism with charge order 
is seen under strong electronic interaction. 
Our results suggest that such charge order can be easily controlled by the magnetic field, 
which possibly give rise to the giant negative magnetoresistance, 
and its relation to phthalocyanine compounds is discussed. 
\end{abstract}
\pacs{71.20.Rv, 71.30.+h, 74.70.Kn}
\maketitle
\newcommand{\Beqn}{\begin{equation}}
\newcommand{\Eeqn}{\end{equation}}
\newcommand{\Beqna}{\begin{eqnarray}}
\newcommand{\Eeqna}{\end{eqnarray}}
\newcommand{\nonn}{\nonumber \\}
\newcommand{\expo}{{\rm e}}
\newcommand{\imag}{{\rm i}}
\newcommand{\nsig}{\bar{\sigma}}
\newcommand{\expct}[1]{\langle #1 \rangle}
\newcommand{\cket}[1]{| #1 \rangle}
\newcommand{\bra}[1]{\langle #1 |}
\newcommand{\ag}[1]{$\langle {\rm #1} \rangle$}
\newcommand{\wtilde}[1]{\widetilde{#1}}
\newcommand{\dagg}{^{\:\dagger}}
\newcommand{\wdagg}{^{\;\dagger}}
\newcommand{\ndagg}{^{\dagger}}
\renewcommand{\Re}{{\rm Re}}
\narrowtext
In the quest of conducting states in molecular solids for the past thirty years, 
charge transfer complexes with the particular 2:1 composition of two different 
kinds of molecules, usually donor and acceptor, have played crucial roles, 
e.g.~TMTSF$_2X$ is the classical example to show 
the first organic superconductivity(SC) and very rich phases on the plane of 
pressure and temperature\cite{TM}. 
The donor molecules have an average valence of +1/2 forming a quarter-filled $\pi$ 
electronic system. 
There, an interesting competition and coexistence 
in the types of insulating states are seen under the strong electronic interaction 
depending on the degree of dimerization; 
charge ordering(CO) and dimer-Mott expected in the absence of (or under weak) dimerization 
and at under strong dimerization, respectively\cite{chemrev}. 
Rich physics in these classes of materials have now been explored. 
In addition, another family including both 
quarter-filled $\pi$-electrons and localized spins showed up. 
For example, in two-dimensional $\lambda$-BETS$_2$FeCl$_4$\cite{kobayashi} 
and one-dimensional(1D) EDT-TTFVO$_2$FeBr$_4$\cite{sugimoto}
each acceptor molecule has $S$=5/2(Fe$^{3+}$), the same population number as electrons, 
whereas in TPP[Fe(Pc)(CN)$_2$]$_2$(phthalocyanine compound) 
each donor molecule has $S$=1/2(Fe$^{3+}$)\cite{tajima}, 
twice the electron number. 
The combination of localized spins with the strongly interacting 
quarter-filled electrons offer even more interesting possibilities like 
field-induced SC\cite{uji}, ferrimagnetism\cite{sugimoto}, etc. 
\par
In this paper we focus on the 1D TPP[Fe(Pc)(CN)$_2$]$_2$ 
and study a simplest model 
(extended Hubbard-Kondo model)
including an antiferromagnetic Kondo coupling, $J$, between the conduction electrons 
and the localized spins, whose Hamiltonian is given by, 
\begin{eqnarray}
\hspace*{-5mm}
{\cal H}\!=-\!\sum_{\expct{ij}\sigma}\!\!t
c_{i\sigma}^\dagger c_{j\sigma}\!+\!\sum_{\expct{ij}}V n_i n_j 
+ \sum_j U n_{j\uparrow}n_{j\downarrow} 
+ \sum_j J \vec{S_j}\!\cdot\!\vec{s_j}.
\label{ham}
\end{eqnarray}
Here, $c_{j\sigma}$, $n_j$, and $\vec{s}_j$ are annihilation, 
number and spin operators at site $j$ for 
conduction electrons, respectively. 
$\vec{S}_j$ represent spin operators of the localized spins with 
$S\!=\!1/2$, and $\expct{i j}$ denote the nearest-neighbor(nn) pair sites. 
We do not take account of the effect of dimerization, 
which holds in TPP[Fe(Pc)(CN)$_2$]$_2$. 
We focus on the quarter-filling of the conduction electrons 
where many organic conductors as well as the topic material reside. 
\par
For the case with $J\!=\!0$, it has been confirmed that the CO insulating(COI) 
state is realized at around $U \geq 4t$ and $V \geq 2t$
\cite{mila,tsuchi}. 
On the other hand, usual Kondo lattice model does not include 
the Coulomb interactions, i.e., $U\!=\!V\!=\!0$.
In this case, when the Kondo coupling 
is larger than some critical value, $J_c(\delta)$, 
dependent of the electron filling, $\delta$\cite{tsune,McCull},
a ferrimagnetic metallic state is realized, i.e.,
all the conduction electrons form singlets with 
the localized spins and the remaining localized spins 
align in one direction.
We call this state as ferromagnetic in the following. 
This model has already provided some pictures of the interplay 
of charge and spin degrees of freedom for the heavy fermion systems\cite{aeppli}. 
Still, the strong electronic interaction adopted in the present model would make the physics richer. 
In this context, we will study the interplay between the COI
and ferromagnetic metal(FM). 
\par
\begin{figure}[tbp]
\psfig{file=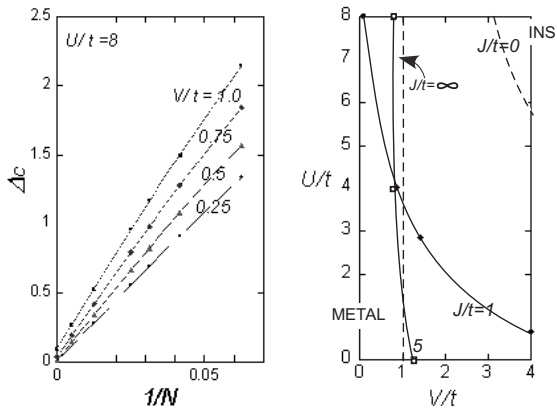,width=7.5cm}
\vspace{-7mm}
\caption{(a) Charge gap $\Delta_c$ as a function of $1/N$ at fixed value of 
$J/t\!=\!1$ and $U/t\!=\!8$ 
for several choices of $V/t$. Typical truncation error is $10^{-6}$-$10^{-7}$ 
with a system size up to $N=200$. 
Symbols at $1/N\!=\!0$ are the extrapolated values, $\Delta_c(\infty )$. 
(b) Ground state phase diagram on the plane of $U/t$ and $V/t$. 
The region above the solid (dashed) lines for $J/t\!=1$ and 5 ($J/t\!=0, \infty$) 
are the COI state where $\Delta_c(\infty)$ remains finite.
}
\label{fig1}
\end{figure}

First we examine whether the COI state of the extended Hubbard 
model ($J\!=\!0$) survives in the finite $J$ region. 
This can be examined by calculating the charge gap, 
$\Delta_c(N)\!=\! E_N(N_e\!+\!2)\!+\!E_N(N_e\!-\!2)\!-\!2E_N(N_e)$, 
and extrapolating it to infinite system size. 
Here, $E_N(N_e)$ is the ground state energy of system size $N$ 
with the electron number $N_e$. 
Figure \ref{fig1}(a) shows some examples of 
the $N$-dependence of $\Delta_c$ 
obtained by the density matrix renormalization group(DMRG) 
method\cite{white}. 
We adopt a so-called non-Abelian algorithm\cite{McCull2,footnote,footnote2} 
with open boundary condition. 
In this algorithm, we deal with only four states per single site, 
$(s, n_e)$=$(1,1),(1/2,0),(1/2,2),(0,1)$, 
classified by the total spin, $s$, and the electron number, $n_e$, 
e.g. (1,1) represents a triplet state formed by a localized 
spin and a conduction electron. 
The use of this reduced basis provides a dramatic performance of calculation 
as well as perfect conservation of the SU(2) symmetry which is quite easily broken 
in the conventional DMRG. 
\par
From Fig.\ref{fig1}(a), the charge gap at 
$N\rightarrow \!\infty$ is estimated by 
fitting the data to $\Delta_c(N)=\Delta_c(\infty)+A/N+B/N^2$, 
where $A$ and $B$ are constants. 
The metal-insulator (MI) phase diagram at several fixed values 
of $J/t$ derived from this extrapolation is shown in Fig.\ 1(b). 
We use the lowest-energy state irrespective of the total spin. 
For $J/t\!=\!0$, Fig.\ 1(b) reproduces the previous results 
by the exact diagonalization in small clusters\cite{mila}, and 
the insulating state is the CO state. 
When $J/t=1$, this COI phase is significantly stabilized. 
This is rather surprising since the charge degrees of freedom 
is easily controlled by a small amount of interaction with 
localized spins. 
By further increasing $J/t$, the phase boundary asymptotically converges 
to the $V/t\!=\!1$ line in the limit of $J/t=\infty$.
\par 
There are several physical reasons of this enhancement of COI 
state by the Kondo coupling.
Basically $J$ introduces a tendency to form a singlet.
(I) Firstly, this reduces the effective hopping of the conduction electrons.
For example, at $J/t \gg 1$, hopping occurs only when one singlet 
is broken at a site and then a new singlet is formed on its nn site.
In this case, the hopping matrix element is 
$t_{\rm eff}=t\times(1/\sqrt{2})^2=t/2$, and thus
$V/t_{\rm eff}$ increases from the original value of 
$V/t$, which favors the CO state.
(II) Another reason is that the singlet formation disfavors the 
double occupancy of the conduction electrons.
This works cooperatively with $U$.
(III) Finally, a singlet state can gain a self-energy by forming a
virtual triplet state on the nn sites.
This occurs only when the nn site is not occupied by another conduction 
electron. This plays a similar role as $V$.
All these effects extend the CO region towards the smaller 
$U/t$ and $V/t$ region.
\par
These phenomena become clearer if we study the strong-coupling region 
of $J/t \!\gg\! 1$ and $U/t \!\gg\! 1$. 
In this region, each single electron forms a singlet with a 
localized spin on the same site. 
We can regard this singlet as a vacant site and the remaining 
unpaired localized 
spin as \lq\lq particle" which cannot have double occupancy. 
Such \lq\lq particle" is denoted by $d_{j\sigma}$ in the following.
The hopping of conduction electrons or singlets is interpreted 
as that of the \lq\lq particles" in the inverse direction. 
For small $t/J$ and $t/U$, we take account of 
the second order perturbation, and the effective Hamiltonian becomes,
\Beqna
{\cal H}_{\rm eff}
&=&\sum_{j \sigma}\Big(\:\frac{t}{2} d\dagg_{j+1\sigma}d_{j\sigma}
 +\frac{t^2}{6J\!+\!4U}d\dagg_{j+2\sigma}d_{j\sigma}(1-n_{j+1}) 
\nonn
& &\hspace{7mm} 
+ \frac{3t^2}{8J}d\dagg_{j+2\sigma}d_{j\sigma}n_{j+1} \Big)  + {\rm h.c.} 
\nonn
&-& \sum_{j \tau \tau'}\frac{t^2}{4J} 
\Big( d\dagg_{j+2\tau}d_{j\tau'} (\sigma)_{\tau\tau'}
   \cdot S^d_{j+1} + {\rm h.c.} \Big)
\nonn
&+& \sum_j \Big( V+ \frac{3t^2}{2J}-\frac{2t^2}{3J+2U} \Big) 
n^d_j n^d_{j+1}  + {\rm const.},
\label{heff}
\Eeqna
where $S^d$ and $n^d$ denote the spin and number operator of 
the \lq\lq particles", respectively. 
The second and the third terms in the first parenthesis give 
modification of the hopping integral, and the next term with $t^2/J$
causes the spin interaction which will be discussed shortly. 
The last term represents the modification of the $V$ term 
as mentioned in the reason (III). 
In the limit of $J\!=\!\infty, U\!=\!\infty$, ${\cal H}_{\rm eff}$ is reduced to 
the $U\!=\!\infty$ limit of the extended Hubbard model 
with $t_{\rm eff}\!=\!t/2$, because the double occupancy of \lq\lq particles''
is not allowed.
Since the nn interaction between \lq\lq particles'' is still equal to 
$V$, the MI phase boundary is $V/t_{\rm eff}=2$, i.e., 
$V/t\!=\!1$, consistent with Fig.~\ref{fig1}(b). 
\par
Let us switch to the magnetic properties. 
Figure 2(a) shows the ground state phase diagram 
classified by the total spin, $S$, obtained in our DMRG. 
The phase transition from 
a paramagnetic to a ferromagnetic state takes place at around 
$J/t\! \sim \!1\!-\!2$ depending on the values of $U/t$ and $V/t$. 
At $U\!=\!V\!=\!0$, a phase transition takes place at $J_c/t\sim 1.6$, in 
concurrence with the results of Ref.\cite{McCull}.
\begin{figure}[tbp]
\centerline{\psfig{file=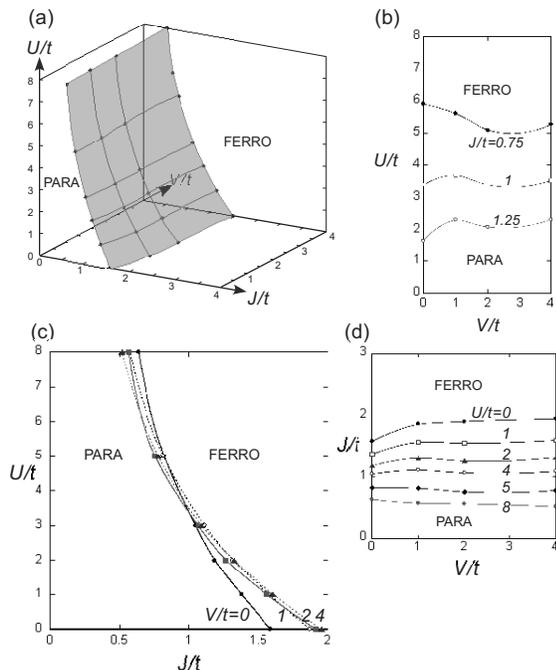,width=7.5cm}}
\vspace*{-5mm}
\caption{Ground-state phase diagrams in the space of (a) $U,V$, and $J$, 
and its cross-section planes (b)$U$-$V$ for several values of $J/t$, 
(c)$J$-$U$ for several values of $V/t$, 
and (d)$V$-$J$ for several values of $U/t$.
Typical truncation error is $10^{-7}$. 
}
\label{fig2}
\vspace{-5mm}
\end{figure} 
The phase transition is the second order in the sense that 
the total $S$ changes successively from 0 to $N/4$. 
However, the region of 
intermediate $S$ is small and depends quite much on the numerical condition 
as in the case with $U\!=\!V\!=\!0$\cite{McCull}, so that 
we determine the phase boundaries from the energy difference between 
$S\!=\!0$ and $S\!=\!N/4$ states. 
We also confirm that 
the spin gap is absent within the numerical error 
at the representative parameter point of each phases. 
Figure~\ref{fig2}(b) shows the ground state phase diagram on the plane of 
$U/t$ and $V/t$ which is to be compared with Fig.~\ref{fig1}(b). 
Apparently, 
the magnetic and MI phase boundaries are not associated with each other
and the spin- and charge-degrees of 
freedom behave independently. 
Although a charge gap opens above some $V_c$, this does not affect 
the spin degrees of freedom much because the charge gap opens 
quite slowly.
\par
The detailed cross-section phase diagram on the plane of 
$U/t$ and $J/t$ is given in Fig.~\ref{fig2}(c). 
The phase boundary moves towards smaller $J/t$ when finite $U/t$ is introduced. 
This is because the singly occupied states of conduction electrons are 
stabilized by $U/t$.
From Figs.~\ref{fig2}(c) and \ref{fig2}(d), we can see that 
the introduction of $V/t$ slightly stabilizes the ferromagnetic phase 
when $U/t$ is large, whereas the opposite tendency is observed 
when $U/t$ is small.
When $V/t$ is larger than $U/t$, two electrons occupy 
a single site rather than to stay apart next to each other, and 
an instability towards superconductivity 
takes place\cite{mila}.
Such double occupancy hinders the ferromagnetism. 
This is, however, quite an unrealistic situation.
\par
In order to study the ferromagnetism in the large $U/t$ region, 
it is convenient to use the effective Hamiltonian of eq.(\ref{heff}).
As in the large-$U$ Hubbard model, the ground state wavefunction 
can be expressed as,
\Beqn
\Psi = \psi_{\rm SF}(t_{\rm eff}, V_{\rm eff})\Phi_{\rm H},
\Eeqn
where a spinless fermion wavefunction, $\psi_{\rm SF}(t_{\rm eff}, V_{\rm eff})$, 
represents the charge degrees of freedom and $\Phi_{\rm H}$
the spin degrees of freedom on a squeezed chain\cite{OgataShiba}.
In the limit of $J=\infty, U=\infty$, 
the spin degrees of freedom are degenerate as in the $U=\infty$ Hubbard
model.
Then the second-order terms in eq.(\ref{heff}) can be considered 
as perturbations\cite{ShibaOgata}. 
The first parenthesis and the last term in eq.(\ref{heff}) 
modifies $t$ and $V$ of the spinless fermions 
into $t_{\rm eff}$ and $V_{\rm eff}$, respectively. 
The degeneracy of the spin degrees of freedom is lifted by the 
remaining terms. 
These terms give an effective ferromagnetic Heisenberg spin 
interaction, $J_{\rm eff}S_i \cdot S_{i+1}$, in the squeezed 
spin chain with
\Beqn
J_{\rm eff} = \frac{t^2}{J}
\langle d\dagg_{j+2}d_{j}n_{j+1} \rangle_{\rm SF} <0,
\Eeqn
where $\langle \cdots \rangle_{\rm SF}$ represents the expectation 
value in $\psi_{\rm SF}$\cite{OgataShiba}. 
This leads to the ferromagnetic ground state of $\Phi_{\rm H}$. 
\par
\begin{figure}[t]
\centerline{
\psfig{file=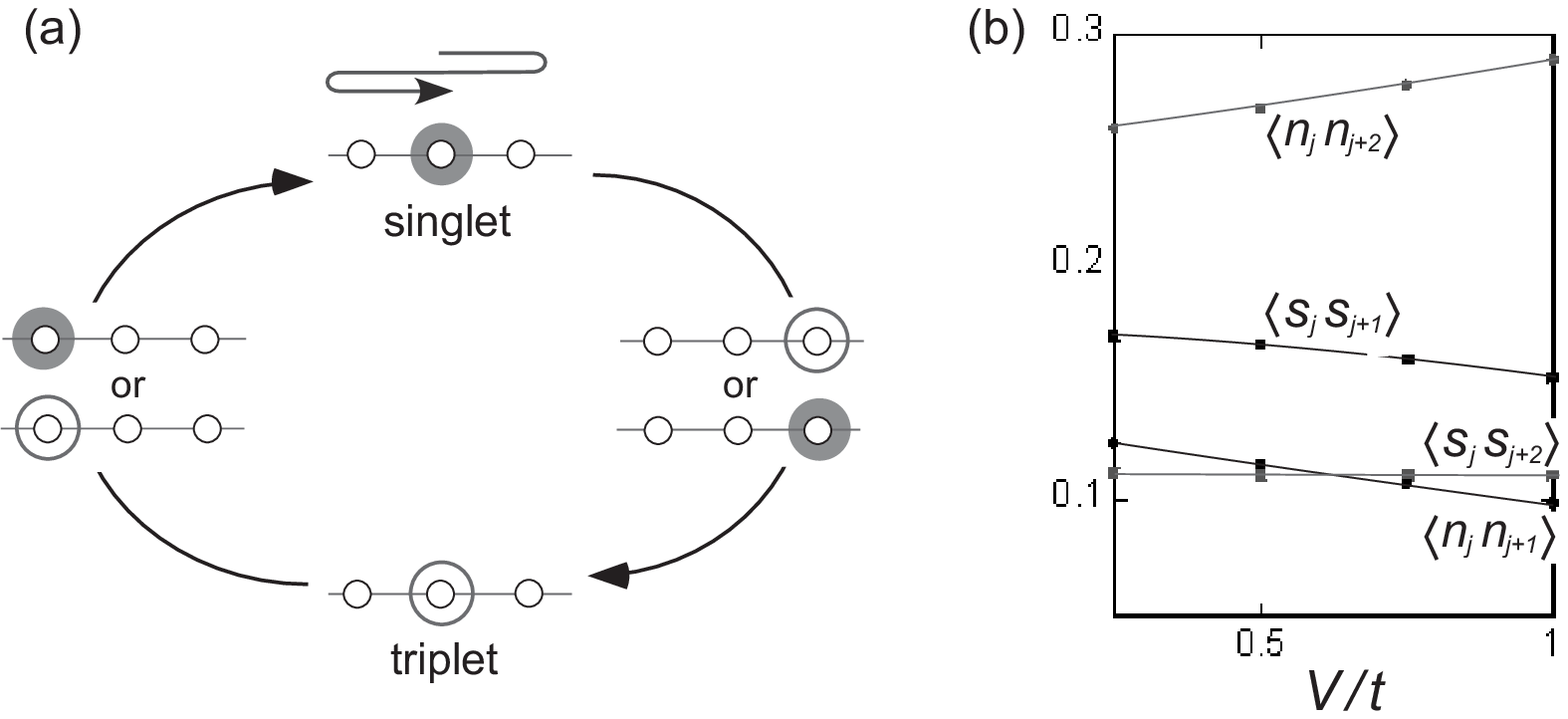,width=7cm}
}
\vspace{-5mm}
\caption{(a) Fourth order perturbation process at $J,U,V \gg t$ in eq.(\protect\ref{ham}). 
Filled and open big circles denote the singlet and the triplet, respectively, 
and the site without them is composed of a free localized spin without electrons.
(b) Localized spin-spin and charge-charge 
correlation functions of nn and nnn sites 
averaged over the whole system as a function of $V/t$ at the fixed value of 
$U/t=8.0$ and $J/t=1.0$. }
\label{fig3}
\centerline{\psfig{file=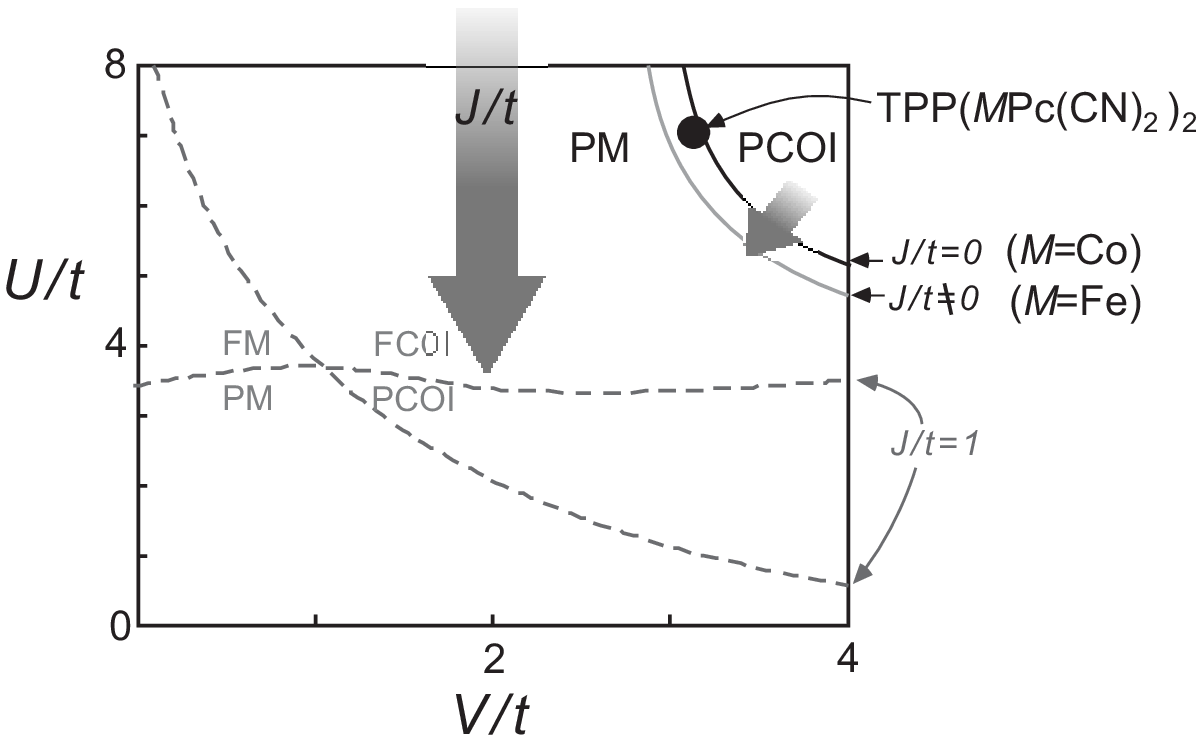,width=6cm}}
\vspace{-4mm}
\caption{Summarized phase diagram on the plane of $U/t$ and $V/t$. Black circle represents 
the possible location of TPP[$M$(Pc)(CN)$_2$]$_2$($M$=Co,Fe).}
\label{fig4}
\vspace{-3mm}
\end{figure}

Let us discuss the $V$-dependence of $J_{\rm eff}$. 
When $V$ or $V_{\rm eff}$ is introduced it reduces 
the nn population of \lq\lq particles", 
which suppresses $J_{\rm eff}$. 
When the CO becomes dominant, we expect a very small $J_{\rm eff}$. 
However, infinitesimal $J_{\rm eff}$ always stabilizes the 
ferromagnetic state because of the degenerate perturbation. 
Actually, we find, 
$\tilde{J}_{\rm eff}= -Jt^4/2V^2(V+J)^2$, 
in the large $V$-region by considering 
the fourth order hopping process of Fig.~\ref{fig3}(a); 
we assume the two localized spins with a singlet in between. 
The electron hops to its nearest neighbor 
to form a singlet or a triplet as a virtual state. 
The triplet formation in the middle of the process 
(after hopping back to the original site) 
allows the mixing between left and right neighboring localized spins to 
give $\tilde{J}_{\rm eff}$. 
Among the four different paths in Fig.~\ref{fig3}(a), the path including singlts 
in the first and third virtual state corresponds to the process included in eq.\ (4).  
Therefore, the ferromagnetism due to 
$J_{\rm eff}$ in eq.\ (4) naturally continues to that of the large $V$-region. 
We calculate the nn and next-nearest neighbor (nnn) 
correlation functions between charges and localized spins 
as shown in Fig.~\ref{fig3}(b); 
nn values are suppressed by $V$ in contrast to 
the nnn ones, which proves the gradual crossover from the ferromagnetism 
by $J_{\rm eff}$ to that by $\tilde{J}_{\rm eff}$. 
\par
\vspace{-4mm}
The ground state phase diagram of eq.(\ref{ham}) is summarized in Fig.~\ref{fig4}; 
at small $J/t$ the phase diagram consists of paramagnetic metal(PM) and 
paramagnetic COI(PCOI) at small and large $U/t$ and $V/t$, respectively. 
With increasing $J/t$, PCOI phase expands. 
At the same time, the ferromagnetic phases (FM and FCOI) descend significantly 
from large $U/t$-part at around $J/t\!\sim\!1\!-\!2$. 
\par
Let us discuss the possible implications of the present theoretical results 
to experiments in TPP[$M$(Pc)(CN)$_2$]$_2$, $M$=Fe,Co. 
In the case of Co-salt which does not have localized spins, i.e. $J$=0, 
a weak sign of CO is observed in the NQR and the resisitivity shows semiconducting temperature dependence 
with a small activation energy ($\Delta_a\!\sim\!10^{-3}$eV). 
This indicates that Co-salts are in the CO region 
but close to the MI phase boundary as shown in Fig.\ref{fig4} in view of $U/t\sim$6-8 
and $V/t\sim$3 with $t\sim$0.1eV given by the extended H\"uckel calculation. 
The magnetic susceptibility behaves as that of the 1D Heisenberg $S$=1/2 
spin system. Here, the antiferromagnetic coupling is expected to be several 
factors larger than the one between spins of completely localized charges 
($J_{\rm AF} = t^4/UV^2\sim$ 10K), 
because of a certain degree of delocalization in the proximity of the MI boundary. 
\\
As for the Fe-salt with localized $S$=1/2 spins, the resistivity is 
semiconducting with larger $\Delta_a \sim 10^{-2}$eV, 
which indicates the stronger CO in Fe salt than in Co-salt. 
This is consistent with the present results that the coupling to the 
localized spins generally enhances the charge gap. 
The magnetic susceptibility in Fe-salt shows large 
anisotropy between those parallel($\chi_{\parallel}$) and perpendicular($\chi_{\perp}$) 
to the 1D chain; 
$\chi_{\parallel}$ behaves similar to the Co-salt case, while $\chi_{\perp}$ obeys the 
Curie-law at $T>T_N\sim$25K and suddenly saturates to become weakly 
$T$-dependent at $T\sim$ 5K-25K, followed by a weak ferromagnetism at $T<$5K. 
Noting that magnetic easy axis of Fe ions is esentilally perpendicular 
to the 1D chain, we may assume that $\chi_{\parallel}$ and $\chi_{\perp}$ are
originated mainly from the ƒÎ electrons and the Fe spins, respectively. 
Hence, the similarity of $\chi_{\parallel}$ with that of 
the Co salt is understood, 
while the behavior of $\chi_{\perp}$ will be interpreted as follows; 
the saturation below $T_N$ suggests that the antiferromagnetic correlation 
developes below $T_N$, which is due to the onset of $\pi$-d interaction 
roughly estimated as $J\sim T_N$. 
In our model, this $J/t\sim$0.025 actually pushes enough the MI phase 
boundary to the region of smaller $U/t$ and $V/t$ to make Fe-salt a PCOI. 
Such small $J/t$ does not lead to 
ferromagnetism which also agrees well with the experiment. 
We speculate that the weak ferromagnetism at $T\sim$5K 
is due to the canted magnetic moments induced by
impurities in the antiferromagnetic background. 
\\
Interestingly, in the Fe-salt, a giant negative magnetoresistance(GNMR) is reported  
below $T\!\sim\!50$K\cite{hanasaki}. 
This will simply be understood in the present framework(eq.(\ref{ham})) 
as due to the suppression of CO phase. 
When the magnetic field is applied, 
the total $S_z$ increases and each site will have the larger population of 
localized $S_z$=1/2 (without electrons) 
as well as of $S_z$=1 triplets composed of one eletron and a localized spin, 
whereas the number of singlets with $S_z$=0 decreases. 
Since the electron hopping from $S_z$=1-site to $S_z$=1/2-site gives $t_{\rm eff}$=1, 
which is larger than those assosiated with $S_z$=0-site($t_{\rm eff}\!=\!t/\sqrt{2}$ or $t/2$), 
$V/t_{\rm eff}$ is reduced and the charge gap will be suppressed. 
This process of GNMR is quite different 
from the conventional double-exchange mechanism\cite{doublex}. 
\\
In conclusion, the CO in the quarter-filled strongly correlated electronic system 
becomes stable when combined with the localized spins, 
and this CO could be easily controled by the magnetic field. 


\begin{references}
\bibitem{TM} D. J\'erome, {\it Science} {\bf 252}, 1509, (2004).
\bibitem{chemrev} H. Seo, C. Hotta, H. Fukuyama, Chem. Rev. {\bf 104}, 5005, (2004), 
and the references therein. 
\bibitem{kobayashi} H. Kobayashi, A. Kobayashi, T. Cassoux, Chem. Soc. Rev., {\bf 29}, 325, (2000).
\bibitem{sugimoto} S. Noguchi, A. Matsumoto, T. Matsumoto, T. Sugimoto, T. Ishida, 
Physica B {\bf 346-347} 397 (2004). 
\bibitem{tajima} T. Inabe, H. Tajima, Chem. Rev. {\bf 104}, 5503 (2004).
\bibitem{uji}S. Uji, H. Shinagawa, C. Terakura, T. Terashima, T. Yakabe, Y. Terai, 
M. Tokumoto, A. Kobayashi, H. Tanaka, and H. Kobayashi, Nature {\bf 410}, 908 (2001).
\bibitem{mila}F. Mila and X. Zotos, Europhys. Lett. {\bf 24}, 133 (1993). 
\bibitem{tsuchi}H. Yoshioka, M. Tsuchiizu and Y. Suzumura, J. Phys. Soc. Jpn. {\bf 69}, 651 (2000).
\bibitem{tsune}H. Tsunetsugu, M. Sigrist, and K. Ueda, Phys. Rev. B {\bf 47}, 8345 (1993).
\bibitem{McCull}I. P. McCulloch, A. Juozapavicius, A. Rosengren, M. Gulacsi, 
Phys. Rev. B {\bf 65}, 052410 (2002). 
\bibitem{aeppli}G. Aeppli, Z. Fisk, Comments Cond. Mat. Phys. {\bf 16}, 155 (1992).
\bibitem{white}S.R. White, Phys. Rev. Lett. {\bf 69}, 2863 (1992). 
\bibitem{McCull2}I. McCulloch, M. Gulacsi, Europhys. Lett. {\bf 57}, 852 (2002). 
\bibitem{footnote}In this algorithm at $S\neq$0, the newly obtained basis after the truncation 
is a linear combination of the old basis with several total spins, 
i.e. the new density matrix has off-block-diagonal elements 
between different total spins. 
However, neglecting the off-block-diagonal elements does not alter the results 
within the numerical error\protect\cite{McCull2}. 
\bibitem{footnote2}
The number of states at each block, $m$, are kept up to 200 which is enough to suppress 
the maximum truncation error to less than the order of $10^{-5}$-$10^{-6}$ in the worst case. 
\bibitem{OgataShiba} M.\ Ogata, H.\ Shiba, Phys.\ Rev.\ B{\bf 41}, 2326 (1990). 
\bibitem{ShibaOgata} H.\ Shiba, M.\ Ogata, Int.\ J.\ Mod.\ Phys.\ B{\bf 5}, 31 (1991). 
\bibitem{hanasaki}N. Hanasaki, H. Tajima, M. Matsuda, T. Naito, T. Inabe, 
Phys. Rev. B {\bf 62}, 5839 (2000).
\bibitem{doublex}A. Moreo, S. Yunoki, E. Dagotto, Science {\bf 283} 2034 (1999); 
K. Kubo, N. Ohata, J. Phys. Soc. Jpn. {\bf 33} 21, (1972). 
\end{references}
\end{document}